\documentclass[pre]{revtex4}
\usepackage[dvips]{graphics}
\usepackage{amsmath}
\newcommand{\degrees}{\hbox{$^\circ$}}
\begin{document}
\title{A nonlinear hydrodynamical approach to granular materials}
\author{Scott A. Hill}
\author{Gene F. Mazenko}
\affiliation{James Franck Institute and Department of Physics, University of Chicago, Chicago, Illinois 60637.}
\date{\today}
\begin{abstract}
We propose a nonlinear hydrodynamical model of granular materials.  We
show how this model describes the formation of a sand pile from a
homogeneous distribution of material under gravity, and then discuss a
simulation of a rotating sandpile which shows, in qualitative
agreement with experiment, a static and dynamic angle of repose.
\end{abstract}
\maketitle

\section{Introduction}

The nature of the theory which describes the macroscopic transport in
granular materials \cite{overview} remains an unsolved problem.  We
propose here a candidate theory, inspired by features of
nonlinear hydrodynamics (NH).  Given its success in
treating transport in a variety of complex systems \cite{NH}, NH
seems natural to use in this context.  In developing any continuum
model of granular materials, however, one is challenged by the need to
incorporate some remnant of the discrete nature of the underlying
material.

The development of our theory is based on the hypothesis that the
states of a system can be specified in terms of a few collective
variables (at sufficiently long length scales), and that these states
are connected in time by the local conservation laws supplemented with
constituitive relations.  Typically, the collective variables used are
the conserved densities which, in the simplest view of granular materials,
are conservation of mass and momentum.  Conservation of energy is
more complicated in this case than in simple fluids, due to locally
inelastic processes \cite{collapse} which may transfer energy into
internal degrees of freedom.  Although we know how to include the
energy density into the description, we begin with a more primitive
theory which uses only the mass density $\rho$ and the momentum density
$\vec g$.

The key difference between simple fluids and granular materials is
that fluids organize over short time scales to be spatially
homogeneous, while granular materials can exist over very long times
in spatially inhomogeneous metastable states: metastable, because
individual grains do not always pack together efficiently;
long-lasting, because the large masses of the grains prevent thermal
fluctuations from adjusting particles into a tighter configuration.
Any complete description of granular materials must be able to explain
the nature of these \emph{quenched} inhomogeneous states.  Experiment
suggests that a sandpile contains ``force chains'' \cite{forcechains}
which serve to support the pile in the presence of external stress;
these chains are surrounded by regions of sand which are comparatively
unstressed, and which allow the pile to flow and to settle under
vibrations.

Since these sandpiles (in the absence of forcing) are static, the
momentum is negligible, and we are left with the density field as our
sole tool in describing the pile.  Our primary hypothesis is that the
density field alone is sufficient to capture the metastable nature of
the system when at rest.  Specifically, we describe the force chains
as regions of sand in which the packing of the grains is ideal, having
the maximum possible density.  We will call these regions
``close-packed''.  The other, ``loose-packed'' regions have their
grains arranged in some non-optimal configuration, with a slightly
smaller density.  We propose that these small variations in the
density field are sufficient to describe the metastable nature of
sand.

Using dynamical equations based on local conservation of mass and
momentum, this \emph{quenched} pattern in the density can be prepared
in a dynamic process driven by an effective free energy with minima
characterizing the loose-packed and close-packed regions.  This
presents the opportunity to grow a sand pile under the influence of
gravity starting from an initial homogeneous state, and in the
construction of the sand pile one builds up, in a self-consistent
manner, an inhomogeneous equation of state characterized by a
nontrivial stress tensor.

The basic structure of our theory, based on local conservation laws,
is straightforward.  More difficult is the inclusion of these
competing close and loose-packed states.  Clearly these require a
detailed description of the shortest length scales treated in the
model; there is a competition of length and energy scales of a type
not encountered in simple fluids.  Such a short-range description will
depend on details of the particular granular material; we will be
satisfied with a theory that demonstrates some generic properties of
sand, as described in the next section.  It seems reasonable that one
may then be able to refine the model to include variations in the
shapes and types of individual grains.

While the short-range length scales give difficulty, the large-scale
nature of our model presents us with the advantage of a description
close in scale to those phenomena which are most visible to experiment
and casual observation.  One ultimate goal is to address the existing
\cite{Knight} macroscopic shaking and rotation experiments.
Also, such a theory should eventually be useful in analyzing the
surface states produced in large-amplitude shaking experiments
\cite{surfacestates}.

There have been a number of attempts \cite{Hydro} to use hydrodynamics
to understand the dynamics of sand piles.  These efforts differ from
the one advanced here in that they do not propose to follow the full
evolution of the system.  There has also been a significant
theoretical effort to describe granular materials through the use of
kinetic theory \cite{kinetictheory}.  This approach is organized at a
more microscopic level, where the connection to the macroscopic,
static, metastable phase of the system is less direct
\cite{foot-kinetic} and coupling to experiment more difficult.  In
many of these cases it is necessary to ``liquefy'' the granular
material in question, giving it so much energy that it loses its
metastable quality which seems to be crucial in describing certain
aspects of a sandpile's behavior.

In this paper we begin by describing some generic features of sand
which we hope to incorporate into our model.  We then proceed to a
development of our dynamical equations from a generalized Langevin
equation.  This will lead into a discussion of the choice of a driving
free energy, which will be crucial in creating the quenched pattern in
the density field.

To test the validity of the theory we turn to simulation on a
two-dimensional lattice.  We begin with a rectangular box containing a
homogeneous distribution of sand under the influence of gravity, and
we show that it does indeed form a sand pile with an inhomogeneous
distribution of loose and close-packed states.  We then turn to
simulations in a circular container, first forming a sand pile using
the same technique, and then rotating the pile at different speeds.
We compare the resulting angles of repose and oscillations about those
angles to corresponding experimental data in the literature.

\section{Properties of Granular Materials}
\label{sec-prop}
We want to construct our model to be compatible with the following
aspects of granular materials:

\begin{enumerate}
\item Granular materials have a clumping property; simulations of
simple systems of inelastically colliding balls provide evidence for
this phenomenon.  One possible explanation for this clumping is the
theory of \emph{inelastic collapse} \cite{collapse}: when two
particles collide inelastically, they lose energy from their
translational degrees of freedom and so recede more slowly than they
approach each other.  On average, the particles stay closer together
than if they had collided elastically, and regions of higher density
build up.  This can also be described in terms of a hydrodynamical
instability \cite{Goldhirsch}.

\label{Pclump}

\item Thermal energies in a sand pile are very small compared to, for
instance, the average gravitational energy.  Thus we should be able to
ignore thermal noise.
\label{Pthermal}

\item Because of the large masses of the grains, one does not expect a
significant vapor pressure above the interface as is found in
liquid-gas systems.  Indeed we expect a \emph{very} dilute gas of
grains above the pile, whose density exhibits a Boltzmann distribution
due to the effect of gravity.

\item Sand piles are strongly driven by gravity and, because of the
lack of thermal effects, the dense sand pile is separated from the
dilute gas above by a sharp interface.\label{Pinterface}

\item Granular materials are strongly disordered, with metastable
structures forming upon creation of a pile.  This is also due to the
lack of thermal effects.  As mentioned in the introduction, there is
evidence of stress chains running through the bulk of sand piles,
which may be associated with the ramified clumps formed in the more
dilute systems studied using kinetic theory.  The important point for
us is that there is competition between somewhat more dense domains
(stress chains and arches) and other, more loosely-packed regions in
the pile. \label{Pmeta}

\item Granular materials are stiff: when moved or jostled the pile can
for a time behave as a solid object.  Thus the sand pile should have a
relatively uniform density.  \label{Pstiff}

\item One specific example of this stiffness is that a sand pile can
maintain a non-horizontal surface.  For example, when a pile of sand
in a container is rotated about a horizontal axis, it does not flow
immediately but waits until its surface passes a certain critical
angle (called the angle of repose) with the horizontal.  According to
Reynolds \cite{Reynolds}, this effect is due to the static
interlocking grains that make up a granular material: the pile must
dilate first, giving these grains freedom to move about, before flow
is possible.

\item It is well known that granular materials can be very sensitive
to boundary and finite-size effects; the famous
\emph{Brazil-nut-to-the-top} phenomenon \cite{Brazil-nut}, for
instance, is strongly influenced by both.
\end{enumerate}
We will touch on these points while developing our model.

\section{Dynamical Equations}
Inspired by nonlinear hydrodynamics, we begin with the generalized
Langevin equation \cite{Ma,Das}
\begin{equation}
\frac{\partial \psi_\alpha(\vec x,t)}{\partial t}
=V_\alpha[\psi(\vec x,t)]-\Gamma_{\alpha\beta}(\vec x)
{\delta F[\psi(\vec x,t)]\over\delta\psi_\beta(\vec x,t)}.
\label{eq-Langevin}
\end{equation}
(Throughout this section, summation is implied over all repeated
indices.)  The variables $\psi_\alpha$ are the ``slow'' fields of
interest in the problem: as mentioned earlier, these are the density
field $\rho(\vec x)$ and the momentum field $\vec g(\vec x)$.  The
first term $V_\alpha[\psi]$ is the \emph{streaming velocity}, and
corresponds to the reversible terms in typical hydrodynamical
equations.  As we will see, it depends on the Poisson brackets of the
slow fields, as well as the derivatives of the Landau-Ginzburg-Wilson
(LGW) effective free energy $F$.  The second term is dissipative in
nature, and $\Gamma_{\alpha\beta}$ is the symmetric matrix of
dissipative coefficients.  Because granular materials are essentiallly
zero-temperature systems (property~\ref{Pthermal}), there is no
thermal noise driving the system.

For our system, the free energy $F$ can be broken up into kinetic,
potential, and external parts:
\begin{equation}F=F_K+F_P+F_E.\end{equation}

From thermodynamics, one has the result that the variable conjugate to
the momentum density is the velocity field:
\begin{equation}
{\delta F\over\delta g_i(\vec x)}\equiv v_i(\vec x)
\equiv{g_i(\vec x)\over\rho(\vec x)},
\label{eq-velocity}
\end{equation}
where $i$ is a vector label.  The kinetic energy contribution to the
free energy thus has all of the momentum dependence:
\begin{equation}
F_K=\int d^d\vec x\,{\vec g^2(\vec x)\over 2\rho(\vec x)},
\end{equation}
while $F_P$ and $F_E$ are functions only of the density.  We will make
further assumptions as to the form of $F_P$ below.  $F_E$ is the free
energy due to external forces: for instance, in the presence of a
uniform gravitational field
\begin{equation}
F_E=\int d^dx\,g\rho(\vec x)(z-z_0),
\end{equation}
where the scalar $g$ is the acceleration due to gravity and $z_0$ is
the bottom of a confining box.

The streaming velocity $V_\alpha$ (in Eq.~(\ref{eq-Langevin})) is
given by the equation
\begin{equation}
V_\alpha[\psi]=\left\{\psi_\alpha,\psi_\beta\right\}
{\delta F\over\delta \psi_\beta}
\end{equation}
where the indices $\alpha,\beta$ run over the set $\left\{\rho,\vec
g\right\}$.  We calculate the Poisson brackets by identifying the
fields $\psi_\alpha$ with microscopic variables, evaluating their
Poisson brackets, and expressing the results in terms of the
$\psi_\alpha$, getting the standard results \cite{Das}
\begin{equation}
\left\{\rho(\vec x),g_i(\vec x')\right\}
=-\nabla_i[\delta(\vec x-\vec x')\rho(\vec x)]
\end{equation}
and
\begin{equation}
\left\{g_i(\vec x),g_j(\vec x')\right\}
=-\nabla_j[\delta(\vec x-\vec x')g_i(\vec x)]
+\nabla'_i[\delta(\vec x-\vec x')g_j(\vec x)].
\end{equation}
With these and the appropriate derivatives of the free energy, we can
calculate the streaming velocities:
\begin{equation}
V_{\rho}=-\nabla\cdot\vec g
\end{equation}
and
\begin{equation}
V_{g_i}=-\rho(\vec x)\nabla_i{\delta F_P\over\delta\rho(\vec x)}
-\rho(\vec x)g\delta_{iz}
-\nabla_j\left({g_i(\vec x)g_j(\vec x)\over\rho(\vec x)}\right).
\end{equation}

By choosing $\Gamma_{\rho\beta}=0$, one easily obtains the usual
continuity equation
\begin{equation}
\frac{\partial \rho}{\partial t}=-\nabla\cdot\vec g.
\end{equation}
The components of the damping tensor associated with the momentum
density,
\begin{equation}
L_{ij}\equiv \Gamma_{g_ig_j},
\end{equation}
can be expressed in terms of the viscosity.  If $L$ were independent
of the fluctuating fields, then we could write it in the general form
\begin{equation}
L_{ij}=-\eta_{il,kj}\nabla_l\nabla_k
\end{equation}
where
\begin{equation}
\eta_{il,kj}=\eta_0(\delta_{il}\delta_{jk}+\delta_{ik}\delta_{jl})
+(\zeta_0-{2\over3}\eta_0)\delta_{ij}\delta_{kl}
\label{eq-eta}
\end{equation}
is the viscous tensor, and $\eta_0$ and $\zeta_0$ are the ``bare''
viscosities.  It is more realistic, however, for the dissipation to
depend on the density of the sand, and so we introduce a function
$\phi(\rho)$ into the damping tensor:
\begin{equation}
L_{ij}=-\eta_{il,kj}\nabla_l\left(\phi(\rho)\nabla_k\right).
\end{equation}
We shall say more about $\phi(\rho)$ below.  It should be emphasized
that, in a system with strong nonlinearities and spatial inhomogeneity
such as ours, there exists no simple relationship between these bare
viscosities and their physical counterparts.

We can now write the remaining Langevin equation:
\begin{eqnarray}
\nonumber\frac{\partial g_i}{\partial t}&=&V_{g_i}
-\Gamma_{g_ig_j}{\delta F\over\delta g_j}\\
&=&-\rho\nabla_i{\delta F_P\over \delta\rho}-\rho g\delta_{iz}
-\nabla_j(g_ig_j/\rho)-L_{ij}\frac{g_j}{\rho}
\end{eqnarray}

At this point, we must choose a form for $F_P$.  In this simple
theory, we make the assumption that $F_P$ is of the square-gradient form
\begin{equation}
F_P=\int d^dx\left[f(\rho)+{1\over2}c(\nabla\rho)^2\right],
\end{equation}
where $f(\rho)$, the free energy density,  is a local function of
$\rho$, and $c$ is a positive constant \cite{foot-sqgradient}.
Thus we have
\begin{equation}
\frac{\partial g_i}{\partial t}
=-\rho\nabla_i\frac{\partial f}{\partial \rho}
+c\rho\nabla_i\nabla^2\rho-\nabla_j(g_ig_j/\rho)
+\eta_{ij,kl}\nabla_j\left[\phi(\rho)\nabla_k(g_l/\rho)\right]
-g\rho\delta_{iz}.
\end{equation}

It should be noted that we can rewrite this equation as the divergence
of a stress tensor plus a term representing external forces:
\begin{equation}
\frac{\partial g_i}{\partial t}=-\nabla_j\sigma_{ij}+F_i^E.
\end{equation}
This is the usual continuity equation for conservation of momentum,
just as our expression for the density's evolution is the continuity
equation for conservation of mass.

For calculational reasons, we prefer to work not with the momenta
$\vec g$, but the velocity fields $\vec v$, defined by
Eq.~(\ref{eq-velocity}) above.  In terms of the density and velocity
fields, the equations of motion become
\begin{equation}
\frac{\partial \rho}{\partial t}=-\nabla\cdot(\rho\vec v)\end{equation}
and
\begin{equation}
\label{eq-dv/dt}
\frac{\partial v_i}{\partial t}=
-\nabla_i\frac{\partial f}{\partial \rho}
+c\nabla_i\nabla^2\rho-v_j\nabla_jv_i
+\frac{1}{\rho}\eta_{ij,kl}\nabla_j\left[\phi(\rho)\nabla_kv_l\right]
-g\delta_{iz}.
\end{equation}
To proceed further, one needs to choose forms for $\phi(\rho)$ and the
viscosities.  For very low densities we expect that the dissipative
part of the equation should vanish; for this to happen, $\phi(\rho)$
must go to zero at least as fast as $\rho^2$.  Higher powers of $\rho$
seem to make our numerical calculations more stable, so we have used
$\phi=\rho^4$.  This issue is not very important in the dense sand
pile where $\rho\approx 1$.  We choose $\zeta_0={5\over3}\eta_0$ in
Eq.~(\ref{eq-eta}) so that $\eta_{ij,kl}$ is isotropic:
\begin{equation}
\eta_{ij,kl}=\eta_0(\delta_{ij}\delta_{kl}+\delta_{ik}\delta_{jl}
+\delta_{il}\delta_{jk})
\end{equation}

In this theory, the input most crucial in specifying sand-like
behavior is the free energy density $f(\rho)$.  With an appropriate
choice of $f$ we will be able to create a system that is very
different from conventional fluids.

\section{Constructing the Free Energy Density}
\label{sec-f}
We will construct the free energy $f(\rho)$ using three steps, as
outlined in Figure~\ref{fig-potentials}.  The mathematical details of
the description of the free energy can be found in the Appendix.

\begin{figure*}[!ht]
\begin{center}
\includegraphics{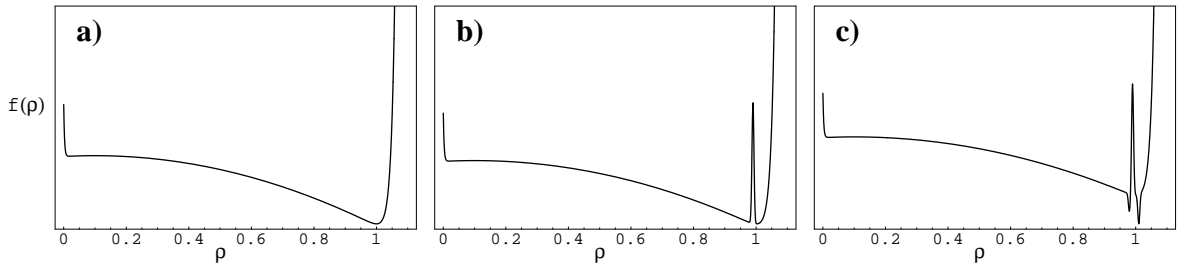}
\end{center}
\caption{\label{fig-potentials} Three steps to our free energy
density: a) with no metastability; b) with a barrier to create
metastability, and c) with wells to stiffen the material.}
\end{figure*}

We can think of $f(\rho)$ as a potential where the system picks out
values of the density which correspond to minima of $f$.  A key
assumption is that, from Property~\ref{Pclump}, grains of sand tend to
clump together, and so the potential has a minimum which we
arbitrarily place at $\rho=1$. The grains themselves are
incompressible, so the potential rises rapidly for densities higher
than $\rho=1$.  Similarly, the potential becomes large as one
approaches zero density, to prevent unrealistic negative values of the
density anywhere in the pile.

Calculations show that this potential, so far, is enough to capture
some properties of sand: the sharp interface of a sand pile, for
instance.  However, it is still missing one key element, and that is
the metastability found in experiment.  Any pile formed using this
free energy would be homogeneous in density, and show no evidence of
the close-packed and loose-packed regions mentioned in
property~\ref{Pmeta}.  Without these regions, our material would be
little more than a slightly compressible liquid.

In Figure~\ref{fig-potentials}b, we introduce this metastability into
the system by placing a barrier into the potential just to the left of
its minimum.  We now have the following picture: the minimum
represents the optimal packing for the sand grains.  As sand comes
together, it increases the density of the pile until it reaches the
density of the barrier where it is frustrated: it can no longer
proceed to the optimal packing.  However, as the sand begins to pile
up, pressures in the pile can provide enough of a push so that some
regions may pass the barrier and reach the minimum of the potential.

This potential has all of the right features, but in simulations it
proves to make a sandpile that is too soft and liquid-like.  To
stiffen the pile, we introduce two wells, as in
Figure~\ref{fig-potentials}c: one for the close-packed states and one
for the loose-packed.  With the wells, this is the free energy we have
used in our calculations.  Future possible modifications to this free
energy will be discussed in our conclusion.

\section{Numerical Analysis}

Because the proposed model has a strongly nonlinear nature, we are
forced to study this system's evolution numerically, hoping that this
will inspire subsequent analytical work.  The first order of business
is to determine numerically whether this system of equations, running
forward in time, can prepare a state that looks like a sand pile.  In
principle, this should be straightforward to do.  In practice, there
are a number of difficulties in putting these partial differential
equations onto a space-time lattice.

\begin{enumerate}
\item {\bf Lattice spacing:} It is understood that, to solve PDE's
numerically, one must keep the time step much smaller than the lattice
size; otherwise, the system develops physical instabilities.  We have
chosen a typical convention of setting $\Delta x=1$ and $\Delta
t=0.001$.

\item We know that there are important {\bf finite-size effects} in
granular systems, and so we must be prepared to examine our model in
various sizes and shapes of containers.  This paper focuses on two
such containers: a rectangle 200 units high and 100 units wide, and a
circle with a 100 unit diameter.  Other containers have been tried as
well, yielding similar results.

\item One of the discouraging aspects of this approach is the large
number of {\bf parameters} governing the system.  There are parameters
to control the intrinsic free energy, the square-gradient energy, as
well as those for external forces like gravity.  There is the finite
size of the box and the nature of the boundary conditions.
Dynamically, there are the viscosities and the initial level of the
density and velocity fields.  In this paper, the values of these
parameters have been chosen to obtain structures that resemble sand
piles.  Most of these choices are reported in the appendix where the
form of the free energy is described in detail; in addition, we take
$c=10$ and $\eta_0=12$.  Further work is needed to map out the range
of parameters for which we obtain physical behavior, and determine
which parameters are ultimately important.

\item This numerical problem would be rather straightforward if not
for a set of persistent unphysical {\bf instabilities}; without care
the system can and will explode in a very unnatural fashion.  These
explosions are of two, closely related types: \emph{runaway
velocities} and \emph{negative densities}.  To alleviate the first, we
have taken the pragmatic step of locally averaging the velocity if it
gets larger than some empirically determined cutoff.

Our free energy density has been designed with a barrier at $\rho=0$,
but negative densities do tend to show up in our calculations,
specifically in the dilute region above the sand pile.  We take two
steps to counter them: when the density of a site is small and
positive ($\rho<0.05$), we average the site's velocity with half of
the neighbors' average velocity.  In addition, when the density dips
below zero, we bring in density from its four neighbors to bring it
above zero.  In treating these instabilities, we have been very
careful not to violate conservation of mass.
\end{enumerate}

\section{Forming a Sand Pile}

As a first example of how our model works, we begin with a nearly
uniform distribution of sand which, under the influence of gravity,
forms a sand pile at the bottom of the container.  Our container is a
two-dimensional box, overlaid with a square lattice 100 units wide and
200 units high.  Boundary conditions on the walls of the container are
non-slip.  Our initial conditions are $\rho=0.5+\Delta\rho$ and $\vec
v=\Delta \vec v$, where $\Delta\rho$ and $\Delta \vec v$ are both
small (0.001 rms) random perturbations in the system.  Because mass is
conserved and because the density inside the formed pile should have a
density about $1$, we expect the interface to form in the middle of
the box, around $z=100$.  In principle, one should average over
ensembles of initial conditions.  That is not necessary for our
purposes here, but will become important in more quantitative work.

\begin{figure*}[!ht]
\begin{center}
\includegraphics{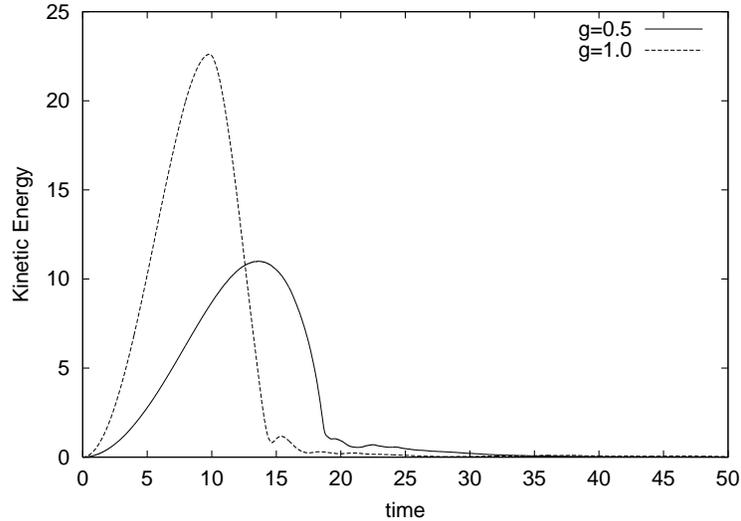}
\end{center}
\caption{Kinetic energies for $g=1$ and $g=0.5$\label{fig-ke}}
\end{figure*}

The first question is whether the model captures the overall dynamics
of the system: our physical picture of the situation has the sand
particles all under the influence of gravity, so at first there will
be a net acceleration.  As particles begin to hit the bottom of the
container, their velocity drops to zero, and the system as a whole
decelerates until it ends up at rest.  To see if this is true we
measure the average kinetic energy
\begin{equation}
KE=\frac{1}{V}\sum_{z,x}\rho(z,x)v^2(z,x)
\end{equation}
where the sum is over the entire box, and $V$ is the total number of
lattice sites.  Figure~\ref{fig-ke} shows a plot of the kinetic energy
for two runs with different gravitational accelerations.

In both cases the behavior is just as expected, with an immediate
acceleration followed by a deceleration to rest.  Also interesting to
note is the difference between the curves: the run with stronger
gravity causes the pile to form faster, with a higher maximum kinetic
energy resulting from the higher initial potential energy of the
system.

To get a clearer picture of the stationary state our system is
settling into, we consider the density profile:
\begin{equation}
P(z)=\frac{1}{N_x}\sum_x \rho(z,x)
\end{equation}
where $N_x$ is the number of sites across the lattice.

\begin{figure*}[!ht]
\begin{center}
\includegraphics{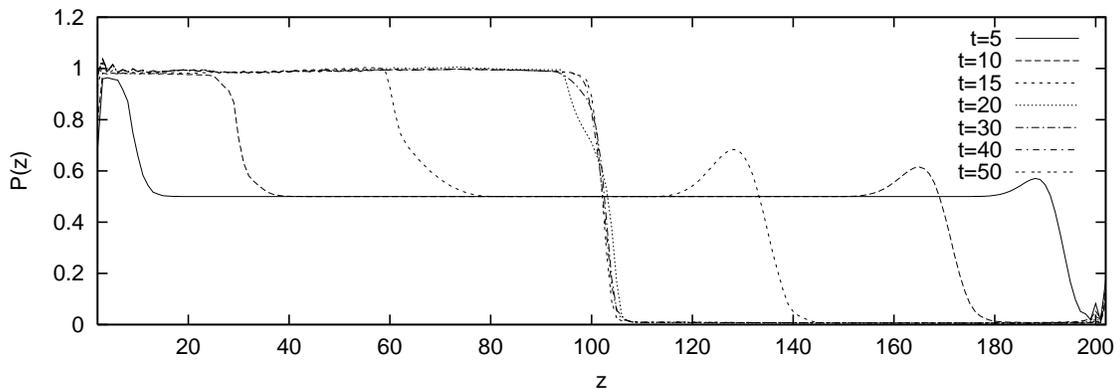}
\end{center}
\caption{\label{fig-prof}The profile of the pile for $g=0.5$,
$t=5,10,15,20,30,40,50$.  The variable $z$ represents the height above
the bottom of the container, which is at $z=0$.}
\end{figure*}

Figure~\ref{fig-prof} superimposes snapshots of this profile at seven
instances in time.  Initially, the profile is a flat line $P(z)=0.5$,
representing the initial homogeneous distribution.  As time
progresses, the density begins to increase for low $z$, indicating
pile formation; at the same time, the density at higher altitudes is
decreasing.  By $t=30$ the system has settled into a phase-separated
system of high-density pile below and low-density ``gas'' above, with
a sharp interface between the two (agreeing with
property~\ref{Pinterface} in section~\ref{sec-prop}.)

\begin{figure*}[!ht]
\begin{center}
\includegraphics{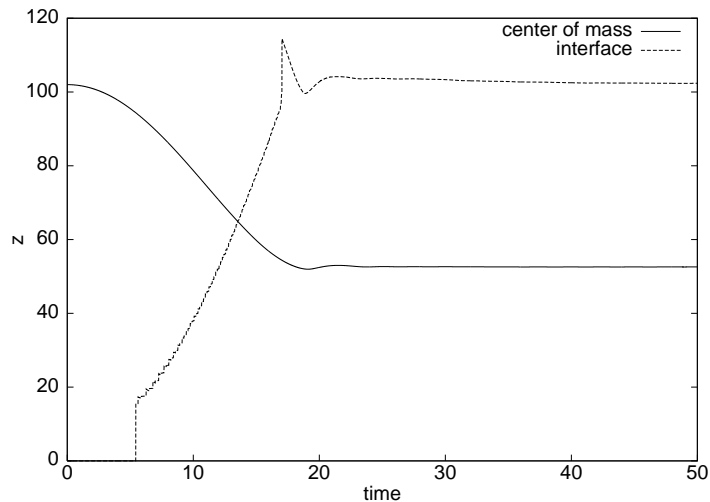}
\end{center}
\caption{The location of the interface, and the vertical component of
the center of mass, for $g=0.5$.  The interface does not really exist
until a pile begins to form, so we did not begin measuring the
location of the interface until some time after $t=5$.  The sharp peak
in the interface's curve occurs when the small bump to the right in
figure~\protect\ref{fig-prof} hits the forming pile below.
\label{fig-interface}}
\end{figure*}

The height of the pile can be defined as that value of $z$ for which
the profile reaches $P(z)=0.5$, or half the average density of the
pile \cite{foot-profile}.  This is plotted as a function of time in
Figure~\ref{fig-interface}, along with the vertical coordinate of the
system's center of mass:
\begin{equation}
z_{CM}={\sum_{x,z} z\rho(z,x)\over\sum_{x,z} \rho(z,x)}
\end{equation}
with the sums over all the points in the lattice.

At first the center of mass moves downward in time parabolically, as
it should with the system in free fall.  As the sand starts to pile
up, less and less of the sand is in motion and the center of mass
approaches a constant height of 52.6.  The height of the interface
moves contrary to this, of course, settling in at 102, about halfway
up the container.  Since the average density of the dense pile should
be about $1$, and we began with a uniform distribution of density
$0.5$, the position of the interface is just a little bit higher than
our expectation that the container be half full of sand, suggesting
that the pile has an average density slightly less than one.  The
center of mass is slightly higher than the midpoint between the base
and the interface of the pile, suggesting that the pile is somewhat
top-heavy, which is surprising.

Since on the large distance scale our system looks like a sand pile,
we turn our attention to the small-scale features inside the pile.
Specifically, we want to investigate the competing regions of
loose-packed and close-packed states, mentioned in
property~\ref{Pmeta} of section~\ref{sec-prop}, corresponding to the
two wells in the free energy.  Recall that these wells are separated
by a barrier, which we have placed at $\rho=0.99$; for definiteness,
we say that all sites with $\rho>0.99$ are close-packed and all sites
with $0.85<\rho\le0.99$ are loose-packed.

\begin{figure*}[!ht]
\begin{center}
\includegraphics{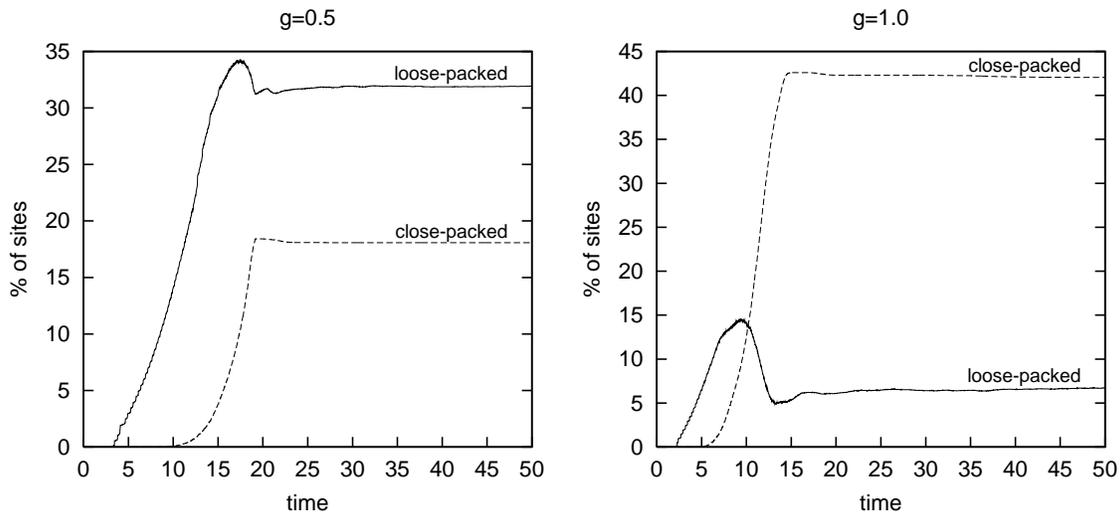}
\end{center}
\caption{Number of loose-packed and close-packed states for $g=0.5$
(left) and $g=1$.\label{fig-stb}}
\end{figure*}

In Figure~\ref{fig-stb} we count the number of loose and close-packed
sites as a function of time, for two different strengths of gravity.
Naturally, the loose sites start to form first; some of these are then
pushed across the barrier and become close-packed due to the weight of
sand on top of them.  Both graphs show the numbers of loose and
close-packed sites approaching a constant as the pile settles.  Notice
that the final ratio of low-density to high-density sites is sensitive
to parameters such as gravity: a stronger gravitational field has a
greater capacity to compact the sand and create more close-packed
sand.

\begin{figure*}[!ht]
\begin{center}
\includegraphics{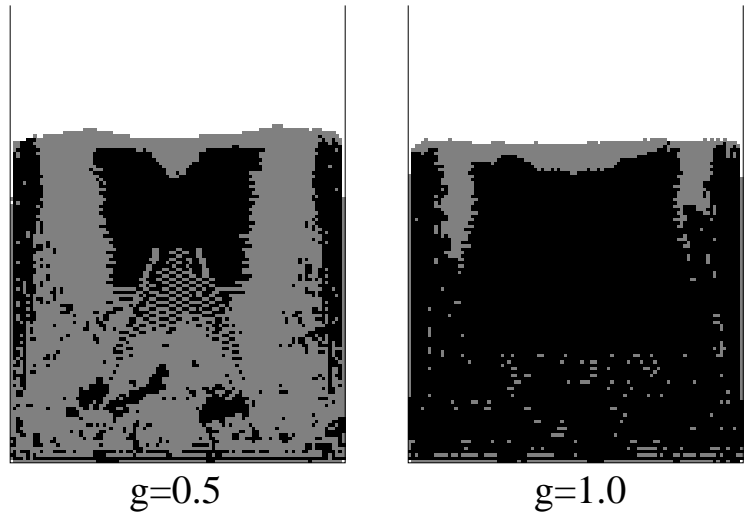}
\end{center}
\caption{\label{fig-pat}Pattern of loose (grey) and close (black)
states for $g=0.5$ (left) and $g=1$.}
\end{figure*}

Figure~\ref{fig-pat} shows how these loose and close-packed regions
are distributed through the pile for the same two values of $g$.  The
top halves of each pile share similar features: both have a layer of
loosely-packed sand on top, where there is no other sand to weigh it
down and compress it.  Immediately below the surface in the center
there is a region of tightly-packed sand, which dips farther below the
surface in the center of the pile than to either side.  This region is
flanked by columns of loosely-packed sand, which are in turn flanked
by dense sand up against the edges of the box.  The differences in the
two piles are mostly at the bottom: the run with less gravity is
mostly loose-packed in the bulk of the pile, while the run with more
gravity is tighter.  Note however that in neither case is the density
uniform: one can find patches of both types in both piles. In
particular, the patterns in the center of the $g=0.5$ pile may hint at
the stress chains that are seen in experiment.

\section{Rotating a Sand Pile}

\subsection{Set-up}
One of the standard ways to probe a granular system is to rotate it
about some horizontal axis.  This method demonstrates several
characteristics of a granular material, and the most important of
these is the angle of repose.  Actually, there are (at least) two such
angles associated with rotation.  If one begins with sand in a
cylindrical container, having a horizontal free surface, and then
begins to rotate the cylinder about its axis, the surface of the sand
will increase in slope along with the container's rotation until it
reaches some critical slope, at which point an avalanche will restore
the surface to a more horizontal state.  This we might call a
``static'' angle of repose, $\theta_s$.  If one continues to rotate
the cylinder, then one of two things will happen:
\begin{enumerate}
\item If the rotation is slow enough, then the flow along the surface
will come to a stop before a further increase in slope should cause an
avalanche again.  The result is a periodic, intermittent flow, and the
interface oscillates about some average angle.
\item If the rotation is faster, the flow from the first avalanche
does not have time to stop before the next avalanche begins.  The
system settles into a state where there is a continuous flow along the
surface, and the interface maintains a constant angle with the
horizontal.
\end{enumerate}
In both cases, we may take the mean angle that the surface makes with
the horizontal over time and call it a ``dynamic'' angle of repose,
$\theta_d$.  In the periodic case, a quantity as important as the
angle of repose is the size of the fluctuation about that angle,
$\delta_d$.  The actual values of these angles seems to depend
experimentally on a number of factors, including particle size and
shape, the humidity of the air, how the pile is formed, boundary
conditions, and so forth.  The experimental results of Jaeger {\it et
al.}  \cite{JLN}, for instance, find that spherical glass beads with
diameters of about half a millimeter show a dynamic angle of repose of
26\degrees, with a fluctuation of 2.6\degrees; while rough
aluminum-oxide particles with the same diameter show a higher angle,
39\degrees, with a fluctuation of 5\degrees.  These seem to be typical
values.  What will be most important in our qualitative analysis is
that these angles exist and are non-negligible.

\begin{figure*}[!ht]
\begin{center}
\includegraphics{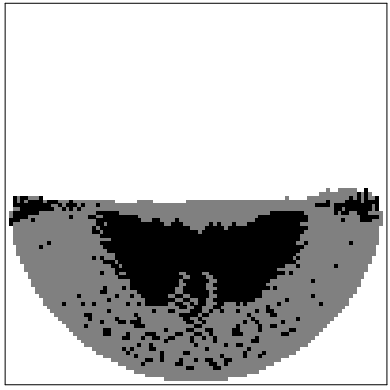}
\end{center}
\caption{\label{fig-circlepat}Pattern of loose (grey) and close
(black) states for $g=1$ in a circular container.  It is interesting
that this looks more like the lower gravity distribution in the
rectangular box, in Figure~\protect\ref{fig-pat}, with a clump of high
density floating on top of a low-density sea.  This is probably due to
the smaller amount of total mass in this system.}
\end{figure*}

We set out to find evidence of these two angles in our model.  To
better match experiments in this field, we move from a rectangular box
to a circular one having a diameter of 100 lattice sites and the same
no-slip boundary conditions.  We grow a sand pile here as we did
before: Figure~\ref{fig-circlepat} shows the pattern of loose and
close-packed states in a pile formed in our circular container.

To best mimic rotation under experimental conditions, one would
ideally set up rotating boundary conditions, giving a constant
velocity to the sand at the edge of the container.  This has turned
out to be difficult to do in practice: one reason is that, due to the
lattice, our boundary is not a perfect circle, and there is a tendency
for mass to ``leak'' in or out of the boundaries when the boundary
conditions are not strictly no-slip.  Because of this, we choose to
implement rotation by rotating gravity with a constant period of
rotation $T$.

To look for an angle of repose we need to measure the angle that the
interface makes with the horizontal (that is, the normal to gravity).
The most direct way to do this is to fit the interface of the pile
with a line, and measure the angle that this line makes with the
horizontal.  This is indeed one of our probes: we define as our
interface those points which are in the pile ($\rho>0.5$) which have a
nearest neighbor outside of the pile, and, to reduce boundary effects,
we throw out those points that are not within one half radius from the
center.

\begin{figure*}[!ht]
\begin{center}
\includegraphics{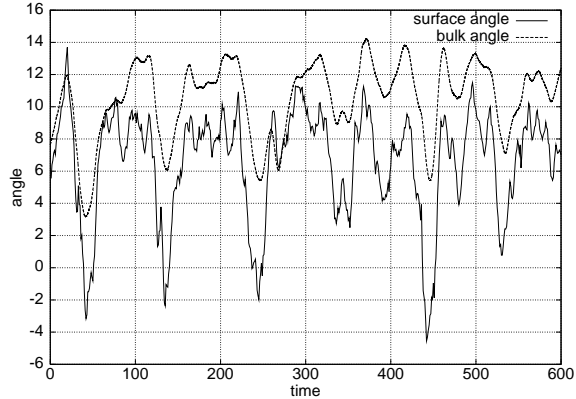}
\end{center}
\caption{\label{fig-compareSandB}The behavior of the surface and bulk
angles for a typical run ($T=200$).  Notice the overall similarity in
the shape of the two plots, although the bulk angle has a much
smoother curve.  The bulk angle is consistently above the surface
angle, suggesting that the pile has an asymmetric shape.  (All angles
are measured in degrees.)}
\end{figure*}

This method should be ideal, but because of the finite size of our
system, this measure ends up depending on only a few points which
makes it sensitive to various local perturbations.  For this reason,
we introduce another measure which we call the ``bulk angle'' (where
the first might be called the ``surface angle'').  This latter measure
is simply the angle that a line passing through the center of the
container and the center of mass makes with the vertical.  This
measure depends on the entire system and so is less sensitive to
noise.  It will be equal to the surface angle in the event that the
pile has reflection symmetry across the vertical axis; it will differ
from the surface angle by some fixed amount if the shape of the pile
during rotation is asymmetric but constant.
Figure~\ref{fig-compareSandB} shows a plot of the two measures for a
typical run.

\subsection{Static Angle of Repose}

\begin{figure*}[!ht]
\begin{center}
\includegraphics{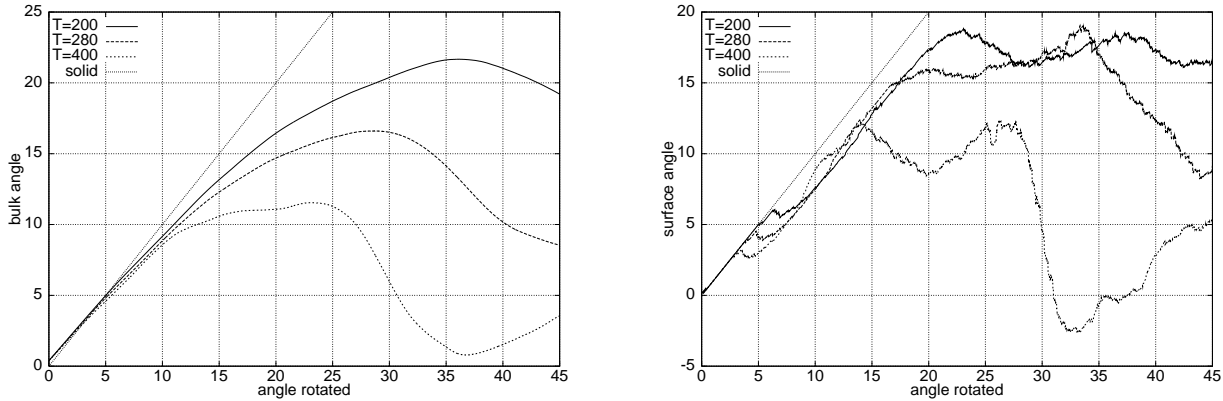}
\end{center}
\caption{\label{fig-initang}The behavior of the bulk (left) and
surface angles during the first 45\degrees\ of rotation, for three
rotation speeds: $T=200$, $280$, and $400$.  The horizontal axis shows
the angle through which the container has been rotated.}
\end{figure*}

With the surface angle and bulk angle measures in place, let us look
systematically at our data.  First, we consider the behavior of the
pile in the first moments of rotation, with three different periods
$T$ (Figure~\ref{fig-initang}).

In all three cases the behavior is similar: as the container turns,
the interface's slope increases, until its angle reaches some maximum
and begins to decline due to one or more avalanches.  Clearly, there
is a static angle of repose here, which depends on the rotation speed.
Notice, however, that if there were no activity in the sandpile before
the first maximum in the angle, then the curve would follow the
straight line denoted in the figure as ``solid'' until turning
downward.  Instead, it seems that the surface is losing slope, lagging
behind the container, from a very early time: there is some
preliminary flow in the pile, even before the first major avalanche.

\begin{figure*}[!ht]
\begin{center}
\includegraphics{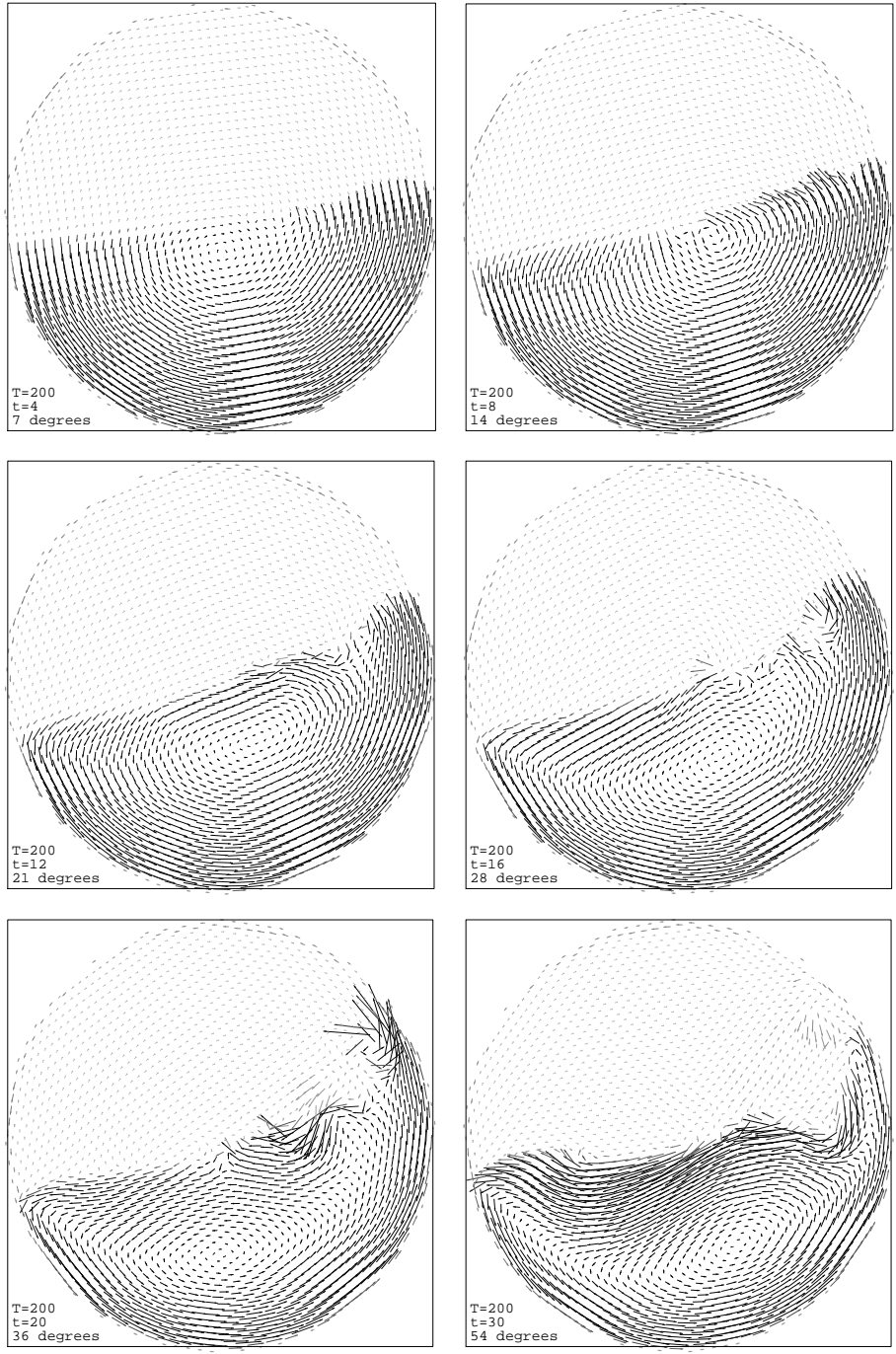}
\end{center}
\caption{\label{fig-initialflows}Still frames depicting the momentum
flow in our pile in the initial stages of rotation, in gravity's frame
of reference.  The darker arrows represent sites in the pile, while
the lighter arrows are the very low-density sites above the pile.  The
length of each arrow is proportional to the momentum, $\rho\vec v$, of
that particular site.  For legibility, each arrow is actually the
average of four lattice sites.}
\end{figure*}

Figure~\ref{fig-initialflows} contains six snapshots of the flow
during this initial period, with $T=200$.  All of the pictures here
are in the lab frame (that is, gravity's frame), so the first picture
($t=4$) shows the pile moving in unison with the container, as if it
were a solid and fastened to the walls.  At $t=8$, however, things are
beginning to change: the center of rotation seems to have moved to the
right and downward, and the sand coming up on the right side is
beginning to curve over to the left.  The result of this action is
seen in the next picture, where there is a definite flow along the
surface of the pile.  Notice that, even though there is surface flow,
it is not enough to prevent the angle of the interface from climbing,
according to Figure~\ref{fig-initang}.  The next picture, at $t=16$,
shows the surface flow continuing, and also that sand is beginning to
climb up the side of the right wall, due to our no-slip boundary
conditions.  At $t=20$, a major avalanche begins, so that the angle of
the interface begins to fall (see Figure~\ref{fig-initang}; this
corresponds to 36\degrees\ on that figure's horizontal axis).  The
last picture ($t=30$) shows a much larger flow of material than seen
earlier.

\begin{figure*}[!ht]
\begin{center}
\includegraphics{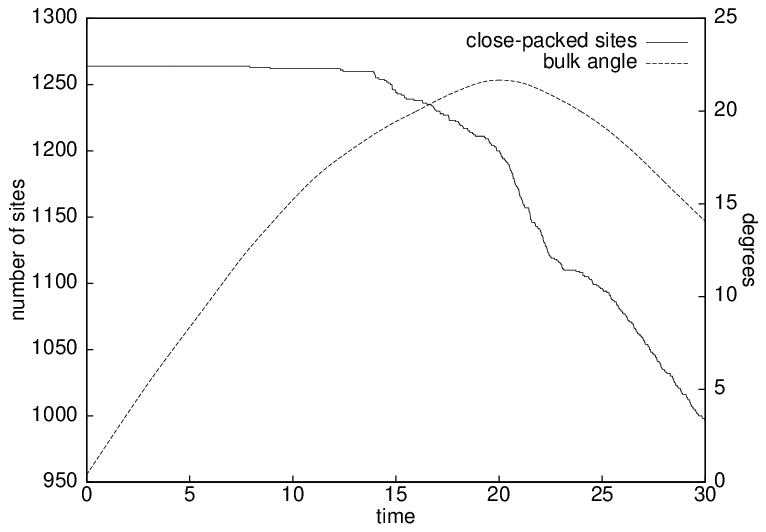}
\end{center}
\caption{\label{fig-closedropoff}The number of close-packed sites for
a cylindrical system rotated at a rate of $T=200$.  Plotted against
this (and on a different $y$-axis) is the bulk angle of the system,
for comparison.  Notice that the bulk angle does not stop its ascent
until after the number of close-packed sites have begun to decline,
and the pile has dilated.}
\end{figure*}

In short, there is clearly an initial period where the sand moves with
the container and where there is no surface flow, evidence for a
nonzero static angle of repose.  Recall from section~\ref{sec-prop}
that this effect is due to the need of the pile to dilate before it
can flow.  In our model, the close-packed sites are the ones which are
restricted in their ability to dilate (because of the barrier in the
free energy).  Thus, if the pile is dilating during this flow, we
expect that the number of close-packed sites in the pile must be
decreasing.  Figure~\ref{fig-closedropoff} shows that this is indeed
so.

\subsection{Dynamic Angle of Repose}

\begin{figure*}[!ht]
\begin{center}
\includegraphics{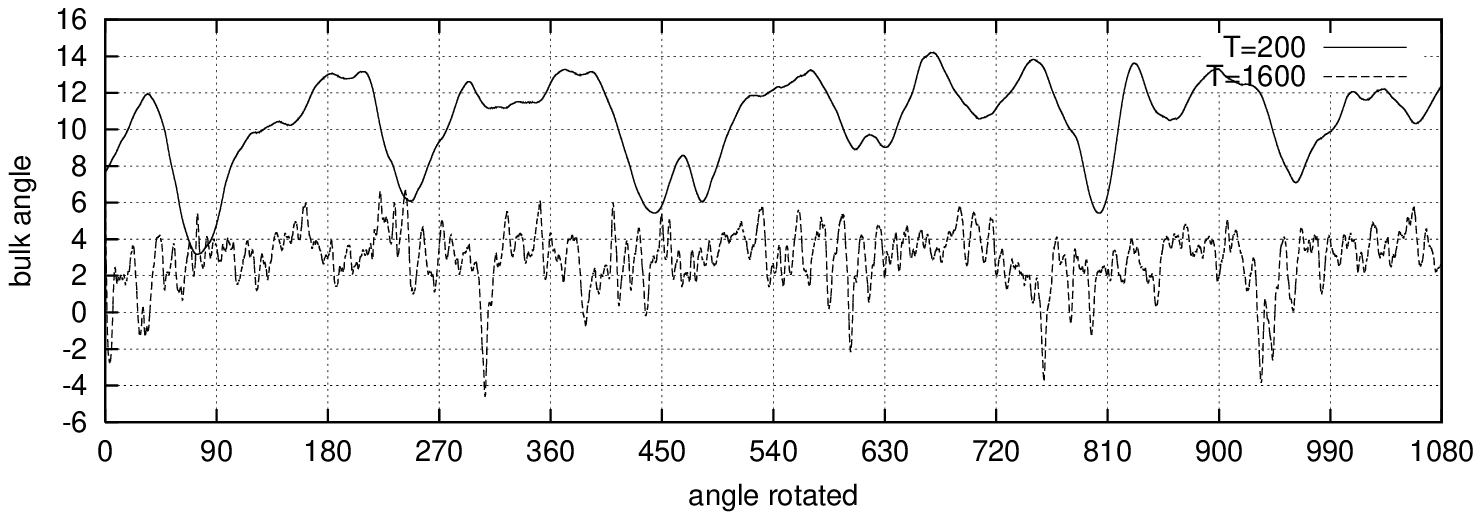}
\end{center}
\caption{\label{fig-cmpbulks}The bulk angle over three revolutions for
two different speeds of rotation.  The horizontal axis is again scaled
to show the angle through which the container has rotated, rather than
the time taken.}
\end{figure*}

We next consider how the pile behaves under further rotation with
three or more turns of the system.  Figure~\ref{fig-compareSandB}
above shows the behavior of the bulk and surface angles over three
complete revolutions, for $T=200$.  Figure~\ref{fig-cmpbulks} below
compares the bulk angles between $T=200$ and $T=1600$ over three
revolutions.

\begin{figure*}[!ht]
\begin{center}
\includegraphics{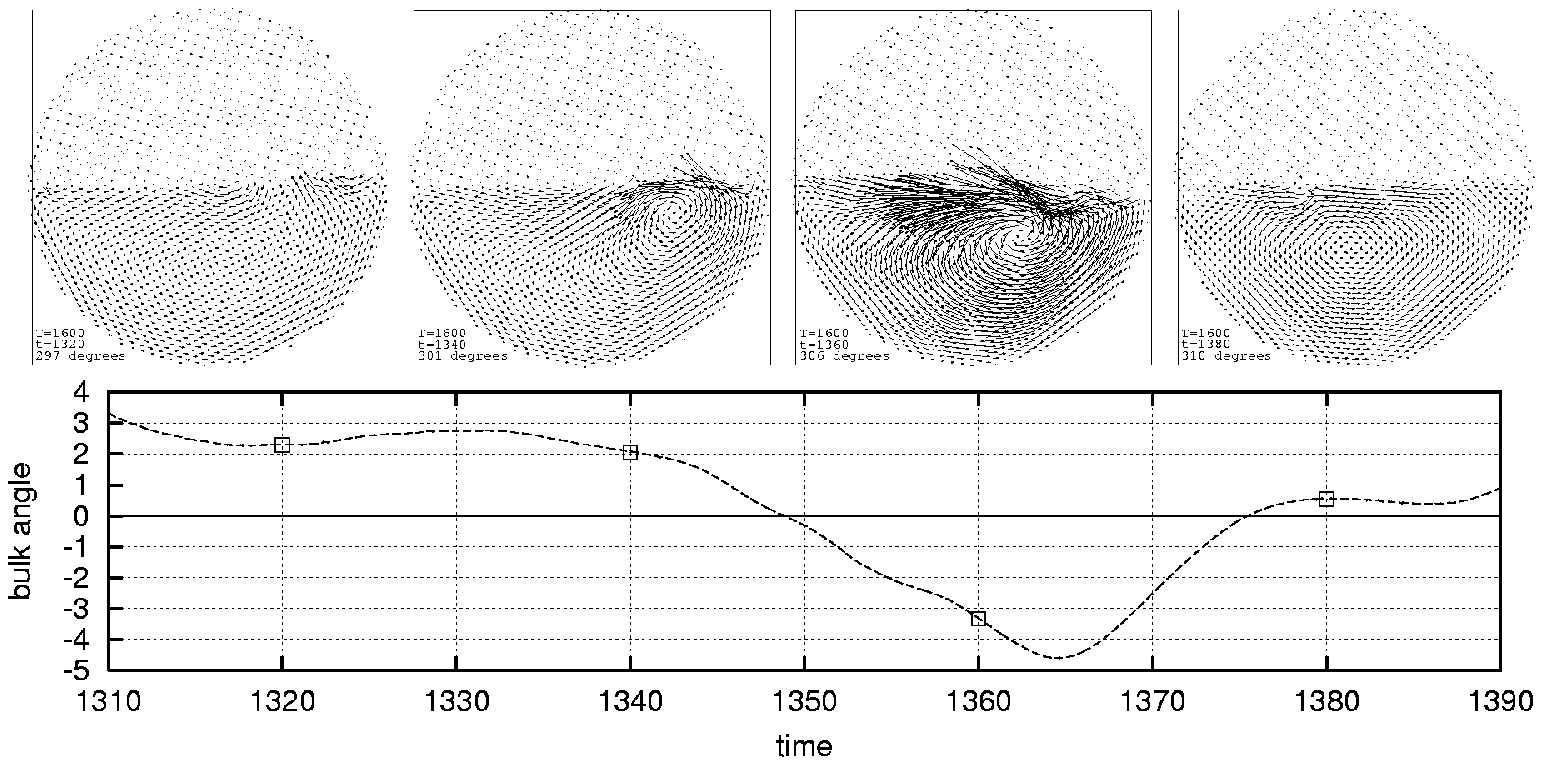}
\end{center}
\caption{\label{fig-negativeangle}An instance in the slow ($T=1600$)
rotation of the pile where the bulk angle dips below zero.  The four
flow diagrams show the momentum of the pile at $t=1320$, $1340$,
$1360$, and $1380$.  The mechanism by which the angle drops is a large
vortex in the pile which give individual grains a large momentum to
the left.  Notice how calm the pile is before and after this event:
certainly this is intermittent behavior.}
\end{figure*}

Most notable in Figure~\ref{fig-cmpbulks} is that the average slope of
the interface is larger for a faster rate of rotation.  Also, the
slower system deviates much less from a constant angle than does the
faster one, with more jaggedness suggesting frequent small avalanches.
These points suggest an inertial effect: in the faster case the sand
does not have as much time to avalanche and level its interface, while
the slower case has a number of short avalanches that help to keep its
interface closer to the horizontal.  However, the slow line is
occasionally punctuated by large avalanches that dip into the
negative: curiously, the interface seems to tilt in the direction
opposite that of rotation every once in a while.
Figure~\ref{fig-negativeangle} looks at one of these dips more
closely.  Apparently, even with the constant small avalanches that we
see in the bulk angle in the slowly rotating case, there is still a
build-up of potential energy that must be released by these larger
events.

\begin{figure*}[!ht]
\begin{center}
\includegraphics{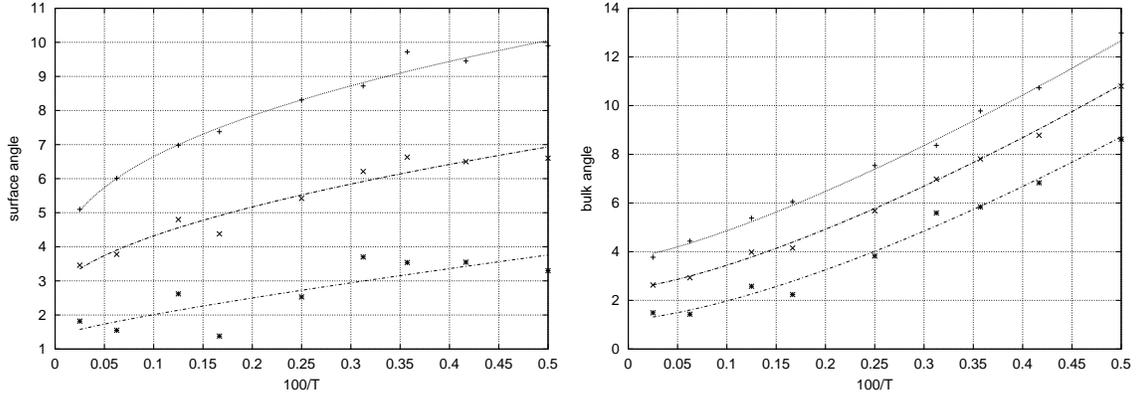}
\end{center}
\caption{\label{fig-dynangs}The average surface (left) and bulk angles
are the center points in each column of their respective plots; the
outer points delineate one standard deviation above and below.  Each
data point represents one run of the system, consisting of the last
two of three complete revolutions; the first turn was thrown out to
diminish initial effects.  The lines are power law fits.}
\end{figure*}

All of our tests so far show a nonzero dynamic angle of repose, but to
rule out the possibility that these angles are due to viscosity or
inertia, it is best to find how the average angle (bulk or surface)
depends on the period of rotation $T$, and from this relationship
extrapolate to $T\to\infty$.  With this in mind we measure the average
bulk and surface angles over time for several rotation speeds, and, in
Figure~\ref{fig-dynangs}, plot them versus the rotation speed $1/T$.

The first thing to note in these plots is that neither angle is going
to zero in the limit of small angular velocities: there is a definite
non-zero angle of repose in our system.  This angle, about
2.5\degrees, is quite small compared to experiment where typically one
finds angles of repose on the order of 30\degrees
\cite{JLN,Rajchenbach}.

\begin{table}
\caption{\label{tbl-fits}We fit the angles of repose in
figure~\protect\ref{fig-dynangs} to the model 
$\theta=\theta_0+c T^\alpha$.}
\begin{center}
\begin{tabular}{lr@{$\pm$}lr@{$\pm$}lr@{$\pm$}lc}
&
\multicolumn{2}{c}{$c$}&
\multicolumn{2}{c}{$\alpha$}&
\multicolumn{2}{c}{$\theta_0$}&
$\chi^2$\\
\hline
bulk angle& 10700&4100& -1.35&0.07& 2.49&0.15& 0.148\\
surface angle& 83.7&105.0&  -0.555&0.275& 2.51&1.12& 0.780\\
\end{tabular}
\end{center}
\end{table}

It is interesting to fit the angles in Figure~\ref{fig-dynangs} to a
power law
\begin{equation}
\theta=\theta_0+c T^\alpha.
\end{equation}
Such fits are shown in the figure, with the parameters given in
Table~\ref{tbl-fits}.  The intercepts $\theta_0$ are positive,
confirming a non-zero angle of repose.  The fit to the surface angle
is not quantitative, but for the average bulk angle the power law fits
well, with an exponent of $-1.4$.

These data points come from single runs, and averaging over an
ensemble of initial conditions may make the fits more quantitative.

Experimentally, Rajchenbach \cite{Rajchenbach} finds that the angular
velocity $\Omega$ goes as
\begin{equation}
\Omega\sim(\theta-\theta_d)^m
\end{equation}
where $m=0.5\pm0.1$.  This corresponds to
\begin{equation}
\theta\sim T^{-2\pm0.4}
\end{equation}
Of our two measures, the bulk angle comes closer to matching
experiment, but there is still some discrepancy that needs to be
addressed.

Finally, to compare the surface and bulk angles with each other.  For
high-speed rotations, the bulk angle is higher on average than the
surface angle, reflecting the plume of material that creeps up the
side of the container (which is accounted for in a bulk calculation,
but specifically excluded from the surface angle).  The surface angle
actually seems to level off at high speeds.  At lower speeds, the bulk
angle actually dips below the surface angle, suggesting that the plume
has disappeared (as is seen in Figure~\ref{fig-negativeangle}).
Notice that the fluctuations in the surface angle are larger than
those in the bulk angle (the former depending on fewer lattice sites
and thus more volatile), and that in both angles these fluctuations
get smaller as one reduces the speed of rotation (but do not seem to
go to zero).

\section{Discussion}

We present evidence that one can create a nonlinear hydrodynamical
model for granular materials that depends only on density and momentum
fields.  Qualitatively, our model demonstrates many key features of
sand, including a sharp interface, a relatively uniform density, a
nonzero angle of repose, and a metastable structure.  The method
allows us to follow the pile from creation forward, in static and
dynamic situations, and the model generally behaves like a sand pile.

There are several ways in which the model could be improved.  The
angles of repose seen here are too small when compared to experiment,
and the way that the angles scale with rotation speed is at odds with
Rajchenbach's findings.  This may not be a problem since we have not
yet looked at the variation of the angle of repose with the parameters
characterizing the model.  An explanation for the difference may also
lie in the fact that our rotation probe differs from the typical
experimental method of rotating a sandpile.  In our simulations, where
we rotate gravity, every particle feels the external force directly
and immediately.  In experiment, where the container is rotated, the
external force must be transmitted inward from the boundaries.  This
difference may be enough to account for the discrepancy between
simulational and experimental outcomes.  One may be able to mimic the
rotation of the container in our model by introducing centrifugal
forces into the system.  Such forces would have a magnitude of
$$F_c=\rho r\left(2\pi\over T\right)^2,$$ where $r$ is the distance from the
center of the container.  Our round container has a diameter of 100
units, so $r\le50$, while $\rho\sim 1$ and the fastest rotation speed
we use is $T=200$; thus these centrifugal forces would have a
magnitude of $0.05$, which is small (though not negligible) compared
to the main acting force of gravity, which is of magnitude roughly
equal to $1$.  Thus centrifugal forces may provide some quantitative
effect, but in our initial, qualitative presentation here we deemed it
unnecessary to include them.

Another way of improving the model is to remove the constraint that
the loose-packed regions have a single fixed density.  There are many
non-optimal, metastable ways to pack particles together, having a
range of different densities.  One solution may be to allow the
position of the loose-packed well to fluctuate slightly at random
through the pile.  There are several plausible ways of implementing
this idea, which we intend to pursue.

We have only begun to extract information using our model; there are
other probes we can use to perturb our system.  Shaking can cause the
pile to slowly settle into a denser state; we hope to find the
logarithmic time dependence seen in compactification experiments
\cite{Knight}.  Applying pressure to the system may allow us to
investigate force propagation in the pile, and determine the nature of
stress chains.  It should be possible to modify the model to depict a
pile made up of two or more types of particles, to investigate the
phenomena of unmixing and the Brazil nut effect in a hydrodynamical
setting.  The flexibility of the fluctuating nonlinear hydrodynamical
approach gives us a wide range of avenues to investigate.

\section*{Appendix: Specification of the Free Energy Density}

The first potential in Figure~\ref{fig-potentials} is made up of four
terms:
\begin{equation}
f_a(\rho)=f_{\text{clump}}(\rho)+f_{\text{large}}(\rho)
+f_{\text{negative}}(\rho)+f_{\text{entropy}}(\rho)
\end{equation}
where
\begin{eqnarray}
f_{\text{clump}}&=&-\frac{1}{2}u_0\rho^2\\
f_{\text{large}}&=&Be^{\lambda(\rho^2-1)}\\
f_{\text{negative}}&=&U_1e^{-\lambda_1\rho}.
\end{eqnarray}
The first term models the clumping behavior of sand mentioned in
property~\ref{Pclump} in section~\ref{sec-prop}.  The second term
models the ultimate incompressibility of the sand grains: $B$ is
chosen so that the minimum of the well is at $\rho=1$.  The third term
prevents the densities from becoming negative by putting a barrier at
zero density.

We have not given the definition for $f_{\text{entropy}}$ yet.  Our
original intent was to have this term model the behavior of the very
dilute gas that exists above our sand pile.  So that the low-density
regions show a Boltzmann distribution, we used the standard gas
entropy term
\begin{equation}
f'_{\text{entropy}}=A(\rho\ln(\rho/\rho_0)-\rho)
\end{equation}
with $\rho_0=0.05$.  However, this was a source of numerical
instability, and since our focus was the pile and the high-density
regions, we replaced this with a simpler term,
\begin{equation}
f_{\text{entropy}}=A\rho.
\end{equation}
The reason we did not eliminate the term entirely was that the
original term created a shallow minimum around $\rho=0.05$, and for
consistency we decided to keep that minimum there which the linear
term allows us to do.

Clearly, this model has a lot of parameters, and one part of our
future work will be to find optimal values for these parameters.  Our
current choices were selected because they gave realistic results: we
set $u_0=2$, $\lambda=40$, $U_1=0.6$, $\lambda_1=400$, and $A=0.2$.
To satisfy the requirement that the minimum of the potential be at
$\rho=1$, we set $B=\frac{1}{2\lambda}(u_0-A)$.

The barrier introduced in Figure~\ref{fig-potentials}b is described by
\begin{equation}
f_{\text{barrier}}=u_be^{-{1\over2}k(\rho-\rho_b)^2},
\end{equation}
where $u_b=1.5$, $\rho_b=0.99$, and $k=10^5$.  The two wells in the
final potential are described by
\begin{equation}
f_{\text{well$_n$}}=-u_ne^{-{1\over2}k(\rho-\rho_n)^2},
\end{equation}
where $\rho_1=0.98$ and $\rho_2=1.01$, $u_1=0.25$ and $u_2=0.4$, and
$k$ is the same as in the expression for the barrier.

\begin{acknowledgments}
We would like to thank Professors Heinrich Jaeger and Sidney Nagel for
helpful conversations.  This work was supported by the Materials
Research Science and Engineering Center through Grant No. NSF DMR
9808595.
\end{acknowledgments}


\begin{thebibliography}{99}
\bibitem{NH}Nonlinear hydrodynamics has been successful in treating
nonlinear phenomena in a wide variety of physical systems ranging from
magnets \cite{lanlif} to superfluids \cite{foot-superfluids}.  The
method has been particularly useful in treating nonlinear behavior
like critical phenomena \cite{Ma}, nucleation theory \cite{Langer},
and unstable growth \cite{Bray}.  It should be appreciated that
nonlinear hydrodynamics is a more complete theory than long wavelength
linearized hydrodynamics.  Conventional hydrodynamics \cite{Martin},
expressed in terms of observable transport coefficients and such,
emerges from NH as a limiting regime.
\bibitem{foot-superfluids}The time-dependent Ginzburg-Landau equation
has played a central role in treating the dynamic properties of
superfluids and superconductors.
\bibitem{foot-kinetic}If one uses kinetic theory, one has the involved
intermediate stage in the theory of deriving hydrodynamics equations
from which one extracts observables.
\bibitem{foot-sqgradient}This form is the simplest we can assume, and
there are a series of more sophisticated choices one can investigate.
\bibitem{foot-profile}Actually, we use $P(z)=0.51$ for computational
reasons.  When no row has exactly this average density, which is often
the case, interpolation is used.
\end{thebibliography}
\end{document}